\documentclass{moriond}
\usepackage{units}
\usepackage{feynmp}
\usepackage{amsmath,bbm}
\usepackage{slashed}
\usepackage{graphicx}
\usepackage{caption}
\usepackage{subcaption}
\usepackage{amssymb}

\newcommand{\de}{\partial}

\renewcommand{\O}{\mathbf{O}}

\renewcommand{\c}{\mathbf{c}}

\newcommand{\beq}{\begin{equation}}
\newcommand{\eeq}{\end{equation}}
\newcommand{\bea}{\begin{eqnarray}}
\newcommand{\eea}{\end{eqnarray}}
\renewcommand{\[}{\begin{equation}}
\renewcommand{\]}{\end{equation}}

\def\be{\begin{equation}}
\def\ee{\end{equation}}
\def\bea{\begin{eqnarray}}
\def\eea{\end{eqnarray}}

\newcommand{\Photo}{\includegraphics[height=35mm]{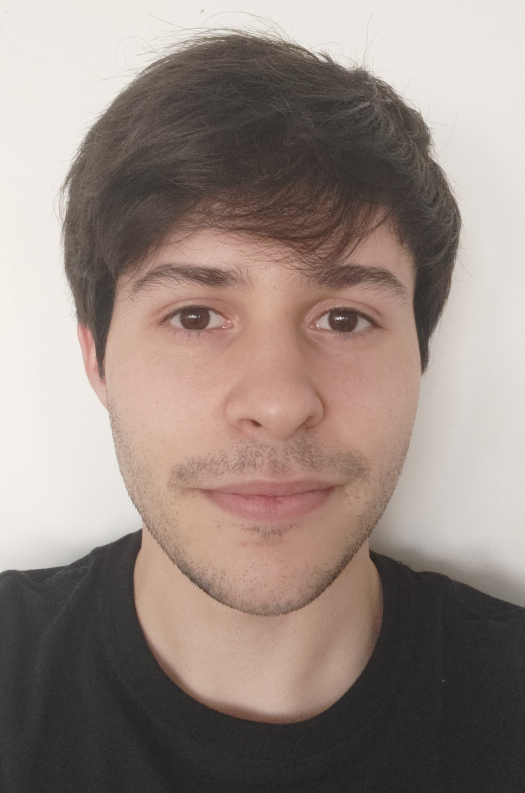}}

\begin{document}
\vspace*{4cm}

\title{ ALP ONE-LOOP CORRECTIONS AND NONRESONANT SEARCHES}
\author{ J.~BONILLA}
\address{Departamento de F\'isica Te\'orica, Universidad Aut\'onoma de Madrid and Instituto de F\'isica Te\'orica IFT-UAM/CSIC, 
Cantoblanco, E-28049, Madrid, Spain}

\maketitle\abstracts{
We consider the  complete leading-order --dimension five-- effective linear Lagrangian for ALPs. The full set of one-loop contributions to ALP-SM effective couplings are derived, including all finite corrections. 
A few phenomenological consequences of these computations are also explored as illustration,  with flavour diagonal channels in the case of fermions: in particular, we explore one-loop constraints on the coupling of the ALP to top quarks. Additionally, we propose a new search for ALPs, targeting VBS processes at the LHC. For this, we consider the tree-level diboson production in VBS, where the ALP participates as an off-shell mediator. Upper limits on ALP EW couplings are obtained from a reinterpretation of Run 2 public CMS VBS analyses. Simple projections for LHC Run 3 and HL-LHC are also calculated, demonstrating the power of future dedicated analyses at ATLAS and CMS.}

\section{Introduction}

Here we summarize the works presented in Refs.~\cite{Bonilla:2021ufe,Bonilla:2022pxu}, in which we explore the phenomenological implications on axion-like-particles (ALPs) couplings: 1) from one-loop corrections and 2) in nonresonant processes at the LHC, respectively.

In particular, in Ref.~\cite{Bonilla:2021ufe} we explore at one-loop order all possible  CP-even operators coupling one pseudoscalar ALP to SM fields at next-to leading order of the linear effective field theory formulation, i.e. mass dimension five operators. The complete one-loop corrections, i.e.  divergent and finite contributions, for an off-shell  ALP and on-shell SM fields is provided and the phenomenological impact is discussed.

In Ref.~\cite{Bonilla:2022pxu} we consider the possibilty of measuring an ALP signal at the LHC. In particular, we explore nonresonant dibson production in vector-boson-scattering (VBS) processes mediated by an ALP, which allow us to explore ALP couplings to EW gauge bosons. New bounds on these couplings are reported by making a reinterpretation of Run 2 CMS analysis on VBS channels.

The results are timely because the level of experimental sensitivity to several ALP-SM couplings has reached a level where 1) one-loop corrections are necessary, and in some cases they are the best tool to constraint some couplings and 2) LHC and other collider experiments can now explore regions of the ALP parameter space which are otherwise inaccessible by other low-energy experiments.

\section{Effective Lagrangian}

A complete  and non-redundant ALP effective Lagrangian is given at $\mathcal{O} (1/f_a)$  by $\mathcal{L}_{ALP} = \mathcal{L}_{SM} + \mathcal{L}_a^{\rm total}$, where $\mathcal{L}_{SM}$ denotes the SM Lagrangian. The most general CP-conserving ALP effective Lagrangian  $\mathcal{L}_a^\text{total}$, including {\it bosonic and fermionic} ALP couplings~\cite{Georgi:1986df,Choi:1986zw}, admits many possible choices of basis.  A complete is that defined  by the Lagrangian
\begin{equation}
 \mathcal{L}_a^\text{total}\,= \frac12 \de_\mu a\de^\mu a + \frac{m_a^2}{2}a^2+\,c_{\tilde{W}}\O_{\tilde{W}}+c_{\tilde{B}}\O_{\tilde{B}}+c_{\tilde{G}}\O_{\tilde{G}}+ \sum_{\text{f}=u,\,d,\,e, \, Q, \, L}\,\text{\bf{c}}_{\text{f}}\, \O_{\text{f}}\,,
\label{general-NLOLag-lin}
\end{equation}
where $\O_{\tilde{X}} = - a/f_a \, X \tilde{X}$ and $\O_\text{f}^{ij} = {\de_\mu a}/{f_a} \, \big(\bar{\text{f}}^i \gamma_\mu  \text{f}^j \big)$. The three coefficients  $c_{\tilde X}$ of the gauge-anomalous operators   are real scalar quantities, while $\c_{\text{f}}$ are 3x3 hermitian matrices in flavour space. In addition, because of the assumption of CP conservation, they obey $\c_{\text{f }}= \c_{\text{f }}^T$. Moreover, in our calculation we neglect flavor mixing (assuming CKM $=\mathbbm{1}$). In this limit, the 6 flavor-diagonal degrees of freedom in the left-handed fermion operators are redundant and are therefore excluded. If CKM mixing is restored, 2 of the diagonal elements for left-handed quarks need to be reintroduced.

\section{Complete one-loop contributions to ALP couplings}

\begin{figure}[t]
\centering
\includegraphics[width=\textwidth]{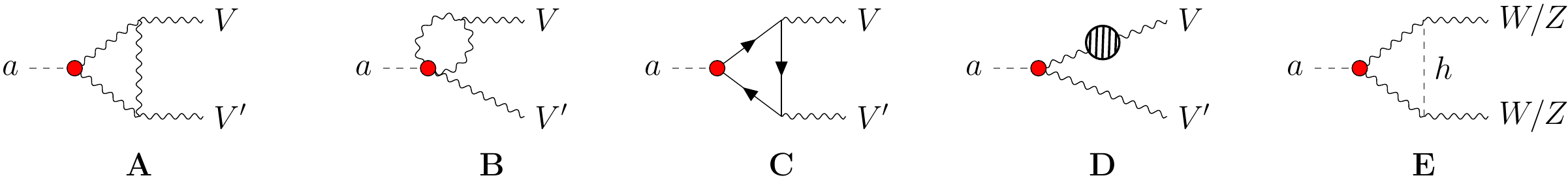}
\includegraphics[width=\textwidth]{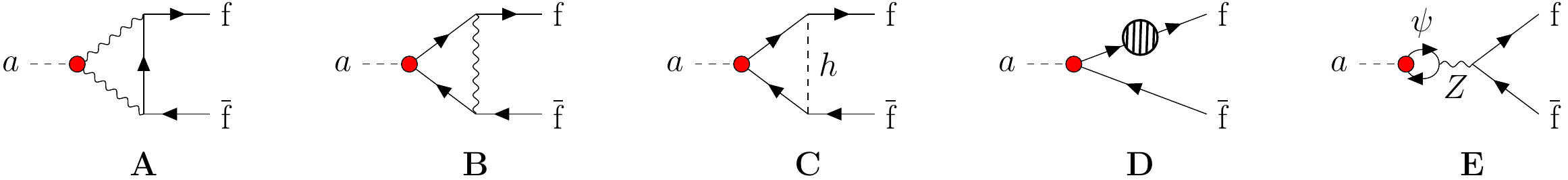}
\caption{One-loop diagrams contributing to bosonic and fermiones ALP interactions at one-loop (the corresponding diagrams with Goldstone bosons and the diagrams exchanging the gauge boson legs are left implicit). $V$ and $V'$ are either a  gluon,  a photon, a $Z$ boson or a $W$ boson. The wavy lines denote all gauge bosons, including gluons. In the last diagram all SM fermions participate in the loop.}
\label{correctionsgammaZ}
\end{figure}

We compute the one-loop contributions to the phenomenological ALP couplings,  including all finite corrections. 1-loop diagrams contributing to bosonic and fermionic couplings are represented in Fig.~\ref{correctionsgammaZ}.

The ALP field is left off-shell (which is of practical interest for collider and other searches away from the ALP resonance), while the external SM fields are considered on-shell. For channels with external fermions, we only provide corrections to the flavour diagonal ones. CKM quark mixing's disregarded in the loop corrections to all couplings. All computations have been carried out in the covariant $R_{\xi}$-gauge, with the help of the {\tt Mathematica} packages {\tt FeynCalc} and {\tt Package-X}~\cite{Shtabovenko:2020gxv,Patel:2016fam}. 

The complete analytic results are lengthy and are not shown here. They are provided in \href{https://notebookarchive.org/2021-07-9otlr9o}{NotebookArchive} in addition with some useful intermediate steps. In the next section we present an example on their use to set new bounds on the ALP parameter space, in particular, to the ALP-top quark interaction.

 \subsection{Some phenomenological consequences: bounds on the ALP-top quark coupling}
 
 We present a example of a situation in which experimental test are able to probe loop corrections: very precise low-energy searches for ALPs which rely on ALP couplings to electrons. 
 
\begin{figure}[t]
\centering
\includegraphics[width=0.7\textwidth]{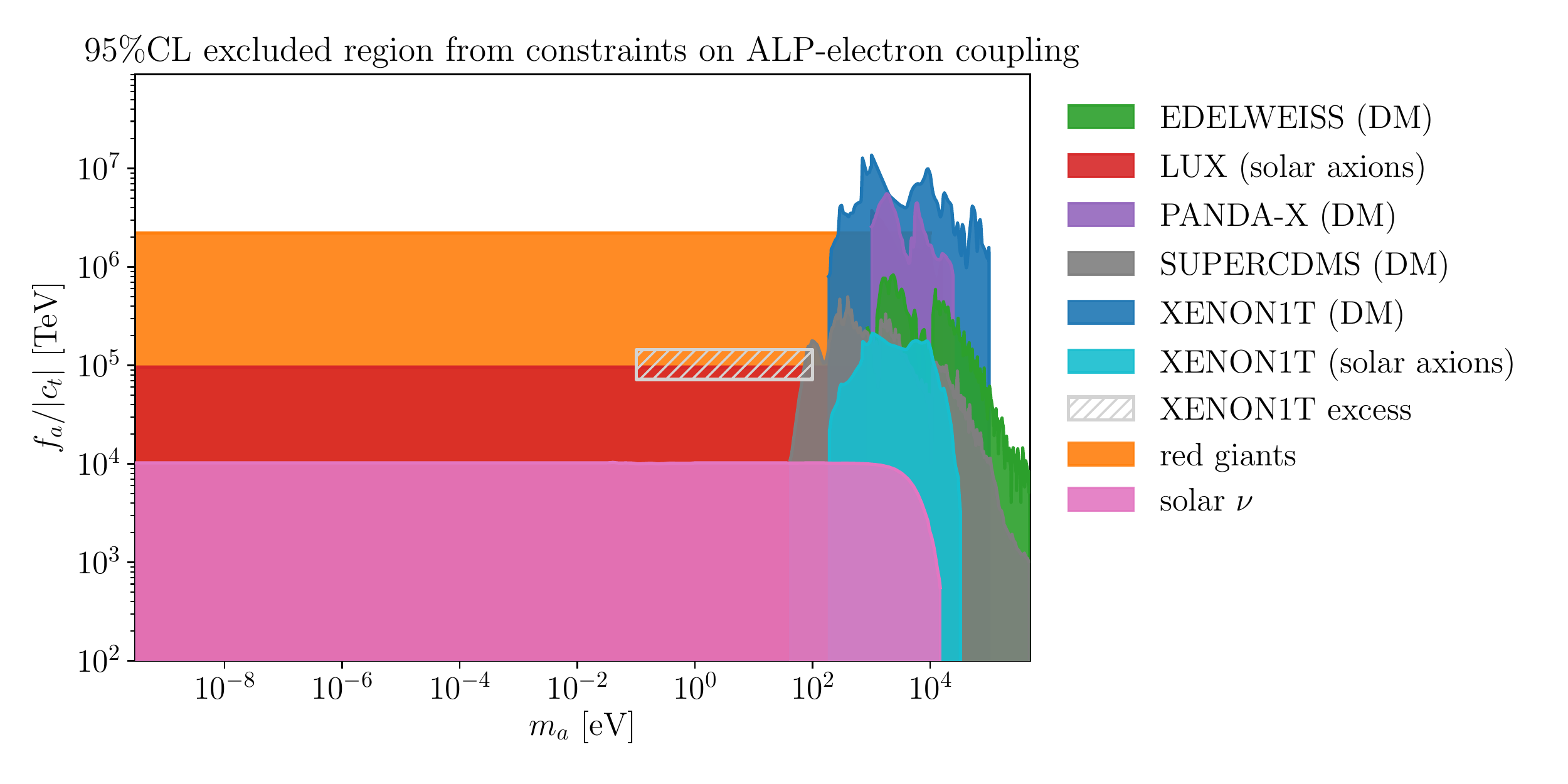}
\caption{Limits on $f_a/|c_t|$ as a function of the ALP mass, extracted rescaling existing constraints on the ALP-electron coupling, taken from Ref.~$^7$.}
\label{fig:bounds_ee}
\end{figure}

 Here we assume that the ALP only couples to top quarks at tree level, so that ALP-electron interactions emerge at one-loop by top-quark mediated processes depicted in the last diagram in Fig.~\ref{correctionsgammaZ}. Then, we consider the current limits on the axion-electron coupling collected in Ref.~\cite{ciaran_o_hare_2020_3932430}, that include results from solar axions searches, ALP DM searches as well as astrophysical bounds. At one-loop level, the ALP-electron interaction is given by (see also Ref.~\cite{Feng:1997tn}):
\begin{equation} \label{cefromctop}
 c_{e}^{\text{eff}} \simeq  2.48 \, c_t \,  \alpha_{em} \,  \log \left( \frac{\Lambda^2}{m_t^2}  \right) \,,
\end{equation}
where we take $\Lambda=\unit[10^6]{TeV}$ in this equation, to extract the bounds on $f_a/c_t$ shown in Fig.~\ref{fig:bounds_ee}. 

\section{Nonresonant searches for ALPs in VBS processes at the LHC}

We have considered the production of a diboson pair in VBS processes mediated by a nonresonant ALP. Typically, ALPs mediating nonresonant processes are assumed to be much lighter than the energy of the process ($ m_a \ll \sqrt{\hat{s}}$), so they cannot be produced resonantly. In this limit, the cross sections happen to be independent of the mass and decay width of the ALP, which allows us to explore vast areas in the ALP parameter space (for a wide range of ALP masses). Typical suppression in the propagator of an off-shell mediator is compensated here due to the derivative nature of the ALP, that leads to an enhancement of the partonic cross sections as: $\sigma \sim \hat{s} / f_a^4$.

The original idea of exploring nonresonant ALP-mediated channels at colliders was proposed in Ref.~\cite{Gavela:2019cmq}, where the authors considered the nonresonant production of a diboson pair in gluon-fusion iniciated processes. In particular, stringent bounds on the product of the ALP-gluon and EW couplings ($c_{\tilde{G}} \times c_{\tilde{V} = \tilde{B} , \, \tilde{W}}$) were derived. 

\begin{figure}[t]
\centering
\includegraphics[width=0.6\textwidth]{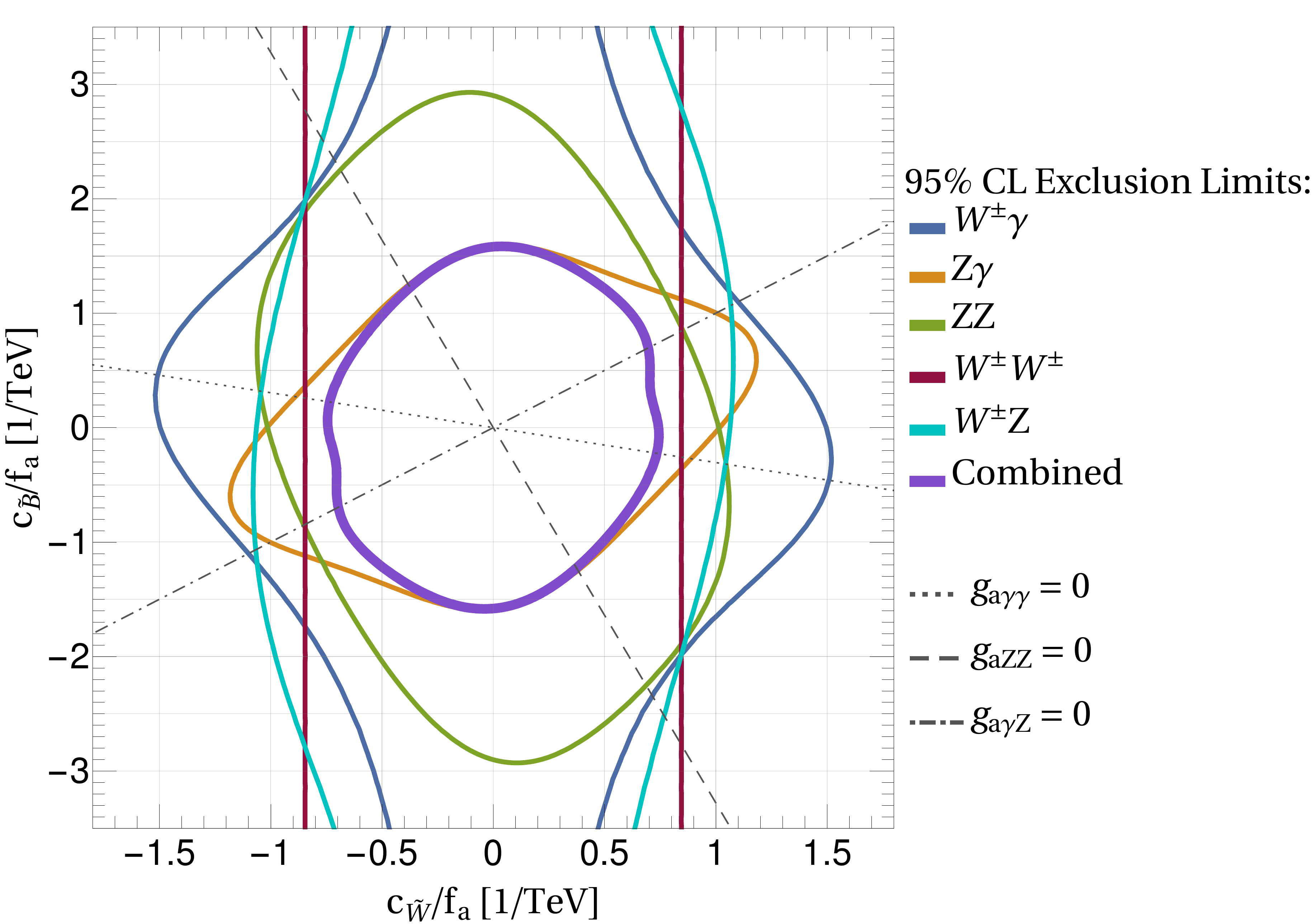}
\caption{Exclusion limits in the $( c_{\tilde{W}} /f_a,\ c_{\tilde{B}} / f_a) $ plane using the data of the Run~2 CMS publications for VBS diboson production channels. The limits have been calculated individually for the five different experimental channels considered and for their combination.}
\label{fig:bounds_VBS}
\end{figure}

In our work we performed a similar analysis but considering the production of dibosons in VBS processes mediated by a nonresonant ALP. As an advantage, these allow us to explore the EW ALP parameters with very reduced dependence on the ALP-gluon coupling, which is of interest for some ALP models in which this coupling is not present (or very suppressed), e.g. the Majoron. By making a reinterpretation of Run 2 CMS measurements of several VBS diboson production channels ($ZZ$, $Z\gamma$, $W^\pm \gamma$, $W^\pm Z$ and same-sign $W^\pm W^\pm$) we have extracted the bounds on $c_{\tilde{B}} / f_a$ and $c_{\tilde{W}} / f_a$ shown on Fig.~\ref{fig:bounds_VBS}. We determined these bounds are valid up to values of $m_a \lesssim$ 100 GeV. Projections for Run 3 and HL-LHC are also computed.

\section{Conclusions}

Here we have summarized the work in Refs.~\cite{Bonilla:2021ufe,Bonilla:2022pxu}. In Ref.~\cite{Bonilla:2021ufe} we computed  the complete one-loop corrections --thus including all divergent and finite terms-- to all possible couplings in the CP-even base for the $d=5$ ALP linear effective Lagrangian, for a generic off-shell ALP and on-shell SM particles.  These results are publicly available at 
\href{https://notebookarchive.org/2021-07-9otlr9o}{NotebookArchive}.

In Ref.~\cite{Bonilla:2022pxu} we propose a new search for ALPs, targeting Vector Boson Scattering (VBS) processes at the LHC. We consider nonresonant ALP-mediated VBS, where the ALP participates as an off-shell mediator. This process occurs whenever the ALP is too light to be produced resonantly, and it takes advantage of the derivative nature of ALP interactions. Upper limits on ALP EW couplings are obtained from a reinterpretation of Run 2 public CMS VBS analyses. Simple projections for LHC Run 3 and HL-LHC are also calculated.

\end{document}